\documentclass[12pt,titlepage]{article}
\usepackage{bm, amssymb,pifont,cancel, amsmath,comment,color,slashed}
\usepackage[dvips]{graphicx}
\makeatletter

\@addtoreset{equation}{section}
\makeatother
 
\begin{document}

\begin{titlepage}
\null
\begin{flushright}
WU-HEP-15-16
\end{flushright}

\vskip 1.5cm
\begin{center}
\baselineskip 0.8cm
{\LARGE \bf Component action of nilpotent multiplet coupled to matter in 4 dimensional ${\cal N}=1$ supergravity}

\lineskip .75em
\vskip 1.5cm

\normalsize

{\large Fuminori Hasegawa} $\!${\def\thefootnote{\fnsymbol{footnote}}\footnote[1]{E-mail address: corni\_in\_f@akane.waseda.jp}}, 
{\large and} {\large Yusuke Yamada} $\!${\def\thefootnote{\fnsymbol{footnote}}\footnote[2]{E-mail address: yuusuke-yamada@asagi.waseda.jp}}

\vskip 1.0em

{\small\it Department of Physics, Waseda University, \\ 
Tokyo 169-8555, Japan}

\vspace{12mm}

{\bf Abstract}\\[5mm]
{\parbox{13cm}{\hspace{5mm} \small
We construct the component action of the system including an ordinary matter and a nilpotent multiplet in global and local supersymmetric framework. The higher dimensional operators of not only Goldstino but also matter and gravitino fields are shown, which appear due to nonlinearly realized supersymmetry. 
}}

\end{center}

\end{titlepage}

\tableofcontents
\vspace{35pt}
\hrule
\section{Introduction}
Supersymmetric (SUSY) theory has been studied as a possible candidate of new physics, a solution for the naturalness problem, and an effective theory of superstring. Phenomenological and cosmological SUSY models have been proposed in many kinds of literature so far. In such models, the SUSY breaking in the early and the present universe is important to realize e.g. inflation and dark energy.

Recently, the SUSY breaking model by introducing a nilpotent chiral multiplet $\hat{X}$, which is called the Volkov-Akulov (VA) multiplet~\cite{Volkov:1973ix,Ivanov:1978mx,Rocek:1978nb}, attracts attention. VA multiplet satisfies the superfield constraint $\hat{X}^2=0$~\cite{Ivanov:1978mx,Rocek:1978nb} and its scalar component $X$ becomes $X\sim G^\alpha G_\alpha/F^X$ where $G_\alpha$ and $F^X$ are the fermionic and auxiliary components of $X$ respectively. Such a multiplet breaks SUSY spontaneously, however, the absence of the scalar component avoids some problems associated with SUSY breaking scalar fields (sGoldstino)~\cite{Antoniadis:2014oya,Ferrara:2014kva,Kallosh:2014via,Aoki:2014pna,Dall'Agata:2014oka,Kallosh:2014hxa,Kallosh:2015lwa,Scalisi:2015qga,Carrasco:2015pla}. For example, the destabilization of a so-called stabilizer during inflation, and also the moduli problem caused by the oscillation of sGoldstino can be avoided in models with a VA multiplet. The VA multiplet is also regarded as an effective action of the anti-D3 brane~\cite{Choi:2005ge,Kallosh:2014wsa,Bergshoeff:2015jxa,Kallosh:2015nia}, which plays an important role in models of moduli stabilization in superstring theory, such as KKLT model~\cite{Kachru:2003aw} and LARGE volume scenario~\cite{Balasubramanian:2005zx}. In the context of the phenomenology, the system with a VA multiplet coupled to minimal supersymmetric standard model (MSSM) has been also studied in Refs.~\cite{Antoniadis:2010hs,Dudas:2011kt,Antoniadis:2011xi,Antoniadis:2012zz,Farakos:2012fm}

The properties of the VA multiplet has also been studied from the theoretical viewpoint~\cite{Komargodski:2009rz,Kuzenko:2011tj}. In Refs~\cite{Casalbuoni:1988xh,Farakos:2013ih}, it is shown that the VA multiplet is an effectively realized in the decoupling limit of the sGoldstino. Constrained complex linear multiplet has also been investigated as a Goldstone multiplet~\cite{Kuzenko:2011ti,Farakos:2015vba,Kuzenko:2015uca}.

In this work, we construct the action of the VA multiplet coupled to a matter multiplet and supergravity (SUGRA). There are some literatures in which the action of the VA multiplet coupled to matter has been studied \cite{Antoniadis:2010hs,Dudas:2011kt,Antoniadis:2011xi,Antoniadis:2012zz,Farakos:2012fm}, however, most of them are models in global SUSY, and the action of a VA multiplet in SUGRA is known only in the SUSY gauge $G_\alpha=0$~\cite{Lindstrom:1979kq,Samuel:1982uh,Samuel:1983ui}, which is not suitable for the case where SUSY is broken by some scalar fields, that is, the Goldstino is a linear combination of the VA fermion $G_\alpha$ and fermionic components of other SUSY breaking multiplets. Therefore, we construct the action in SUGRA without any SUSY gauge conditions, which would be useful for e.g. an investigation of the reheating and the gravitino production in inflation models with the VA multiplet. As we will show, in the case without any SUSY gauge conditions, higher order interactions associated with the VA fermion appear, which may affect the high scale phenomena, such as reheating processes in the early universe.

The remaining parts of this paper are as follows. First, in Sec.~2, we show a systematic way to construct the action with the VA multiplet, which is somewhat complicated because of the unusual structure of equation of motion (E.O.M) of the auxiliary field $F^X$. As demonstrations of our method, we show the VA action with and without a matter multiplet in global SUSY case and find that the VA action is reproduced. Then, we construct the VA multiplet action coupled to SUGRA and a matter multiplet in Sec.~3. Finally, we conclude in Sec.~4. 
\section{VA multiplet action in global SUSY}\label{global}
\subsection{Case without matter multiplets}
First, we discuss the action of the VA multiplet as a demonstration of our systematic way to construct it. In this section, we use the notation in Ref.~\cite{Wess:1992cp}. The VA multiplet $\hat{X}=X(y)+\sqrt{2}\theta G(y)+\theta \theta F^X(y)$, where $y^a=x^a+i\theta\sigma^a\bar{\theta}$ and $a$ is the index of the Minkowski spacetime, satisfies a superfield constraint
\begin{align}
\hat{X}^2=0.
\end{align}
The constraint can be solved in terms of the scalar component $X$ and we obtain
\begin{align}
X=\frac{G^2}{2F^X}.\label{const}
\end{align}
The action we discuss is
\begin{align}
\mathcal{L}=&\int d^4 \theta \hat{X}\bar{\hat{X}}+\left(\int d^2\theta f \hat{X}+{\rm h.c.} \right)\nonumber\\
=&i\partial_a\bar{G}\bar{\sigma}^aG+F^X\bar{F}^{\bar{X}}-\partial_a\bar{X}\partial^aX+(fF^X+{\rm h.c.})\nonumber\\
=&i\partial_a\bar{G}\bar{\sigma}^aG+F^X\bar{F}^{\bar{X}}-\partial_a\left(\frac{\bar{G}^2}{2\bar{F}^{\bar{X}}}\right)\partial^a\left(\frac{G^2}{2F^X}\right)+(fF^X+{\rm h.c.})\label{glVA}
\end{align}
where $f$ is a complex constant.\footnote{The Lagrangian~(\ref{glVA}), which is realized with $K=|\hat{X}|^2$ and $W=f\hat{X}$, describes a general system with a single nilpotent superfield in global SUSY, because $\hat{X}^2=0$ and other possible K\"ahler and superpotential terms given by $\Delta K= \alpha +\beta(X+\bar{X})$ and $\Delta W=\gamma$, where $\alpha,\beta,\gamma$ are constants, do not contribute to Lagrangian. } To obtain the on-shell action, we have to solve the E.O.M of $F^X$, which can be obtained by the variation of $\bar{F}^{\bar{X}}$;
\begin{align}
F^X-\frac{\bar{G}^2}{2(\bar{F}^{\bar{X}})^2}\partial^2 \left(\frac{G^2}{2F^X}\right)+\bar{f}=0.\label{glEOM}
\end{align}
This equation is nonlinear with respect to $F^X$ and $\bar{F}^{\bar{X}}$, and then it is difficult to solve the equation. However, without loss of generality, we can assume that the solution of the equation takes a form
\begin{align}
F^X=A+BG^2+C\bar{G}^2+DG^2\bar{G}^2,\label{ansatz}
\end{align}
where $A,\ B,\ C$ and D are the complex function of quantities other than $G^2$ and $\bar{G}^2$. The ansatz can be confirmed by the fact that $G_\alpha$ and $\bar{G}_{\dot{\alpha}}$ are the Grassmann number, and the indices $\alpha$ and $\dot{\alpha}$ run over only $1$ and $2$. Then, e.g. $G^3$, $\bar{G}^3$ vanish identically. Thus, the ansatz~(\ref{ansatz}) is a general expression of the solution. By substituting the ansatz~(\ref{ansatz}) into the E.O.M~(\ref{glEOM}), we obtain the following equation,
\begin{align}
&A+BG^2+C\bar{G}^2+DG^2\bar{G}^2\nonumber\\
=& -\bar{f}+\frac{\partial^2 G^2}{4\bar{A}|A|^2}\bar{G}^2-\left(\frac{C\partial^2\bar{G}^2}{4|A|^4}+\frac{\bar{C}\partial^2G^2}{2\bar{A}^2|A|^2}\right)G^2\bar{G}^2\nonumber\\
&-\frac{\partial_a(G^2)\partial^aA}{2|A|^4}\bar{G}^2-\frac{\Box A}{4|A|^4}G^2\bar{G}^2+\frac{\partial_aA\partial^aA}{2|A|^4A}G^2\bar{G}^2,
\end{align}
Comparing the coefficients of the series of $G^2$ and $\bar{G}^2$ on the left-hand side with one of the right-hand side, we obtain the following set of the solution,
\begin{align}
A=-\bar{f},\quad B=0,\quad C=-\frac{\bar{f}}{4|f|^4}\partial^2G^2,\quad D=\frac{3\bar{f}}{16|f|^8}\partial^2G^2\partial^2\bar{G}^2,
\end{align}
and, therefore, we obtain the exact solution of $F^X$ as
\begin{align}
F^X=-\bar{f}\left(1+\frac{1}{4|f|^2}\bar{G}^2\partial^2G^2-\frac{3}{16|f|^8}G^2\bar{G}^2\partial^2G^2\partial^2\bar{G}^2\right).
\end{align}
Using the solution, we obtain the on-shell Lagrangian in Eq.~(\ref{glVA}),
\begin{align}
\mathcal{L}=i\partial_a\bar{G}\bar{\sigma}^aG-|f|^2+\frac{1}{4|f|^2}\bar{G}^2\partial^2G^2-\frac{1}{16|f|^6}G^2\bar{G}^2\partial^2G^2\partial^2\bar{G}^2,\label{glVAac}
\end{align}
which is the same with the result in Ref.~\cite{Komargodski:2009rz}, which is equivalent to the original VA action~\cite{Volkov:1973ix} as shown in Ref.~\cite{Kuzenko:2010ef}.
\subsection{Case with a matter multiplet}
We extend the VA action given by Eq.~(\ref{glVAac}) to the one coupled to matter multiplet $\Phi=\phi(y)+\sqrt2\theta\chi(y)+\theta\theta F^\phi(y) $;
\begin{align}
{\mathcal L}=&\int d^4\theta (\hat{X}\bar{\hat{X}}+\Phi\bar{\Phi})+\left[\int d^2\theta (f(\Phi)\hat{X}+g(\Phi))+h.c.\right] \nonumber \\
=&-\partial_{a}\phi\partial^{a}\bar{\phi}-\partial_{a}\left(\frac{G^2}{2F^X}\right)\partial^a\left(\frac{\bar{G}^2}{2\bar{F}^{\bar X}}\right)+i\partial_a\bar{\chi}\bar{\sigma}^a \chi+i\partial_a\bar G\bar{\sigma}^a G\nonumber \\ 
&-\left\{\frac{1}{2}(f''X+g'')\chi^2+f'G\chi+h.c.\right\}\nonumber\\
&+F^\phi\bar{F}^{\bar\phi}+F^X\bar{F}^{\bar X}-\{F^Xf+F^\phi(f'X+g')+h.c.\},\label{mcVAac}
\end{align}
where $f(\Phi)$ and $g(\Phi)$ are holomorphic functions of $\Phi$, and the primes on $f$ and $g$ denote the derivative with respect to $\phi$.
In this case, for a technical reason, we first solve the E.O.M of $F^\phi$ before solving that of $F^X$. Taking a variation for $\bar{F}^{\bar\phi}$, we obtain
\begin{align}
{F}^\phi=\bar{f}'\bar{X}+\bar{g}'
\end{align}
and substituting the solution into Eq.~(\ref{mcVAac}), simplified Lagrangian is written as follow
\begin{align}
{\mathcal L}=&-\partial_{a}\phi\partial^{a}\bar{\phi}
-\partial_{a}\left(\frac{G^2}{2F^X}\right)\partial^{a}\left(\frac{\bar{G}^2}{2\bar{F}^{\bar X}}\right)+i\partial_a\bar{\chi}\bar{\sigma}^a \chi+i\partial_a\bar G\bar{\sigma}^a G \nonumber\\
&-\left\{\frac{1}{2}(f''X+g'')\chi^2+f'G\chi+h.c.\right\}-|f'X+g'|^2\nonumber\\
&+F^X\bar{F}^{\bar X}+(F^Xf+h.c.).\label{mcVAac2}
\end{align}
To obtain the on-shell action, we next have to solve the E.O.M of $F^X$ which can be obtained by a variation of $\bar{F}^{\bar{X}}$;
\begin{align}
F^X-\frac{\bar{G}^2}{2(\bar{F}^{\bar X})^2}\left[\partial^2\left(\frac{G^2}{2F^X}\right)-\frac{1}{2}\bar{f}''\bar{\chi}^2-\bar{f}'g'\right]+\bar{f}-\frac{G^2\bar{G}^2}{4|F^X|^2}\frac{|f'|^2}{\bar{F}^{\bar X}}=0.\label{mcEOM}
\end{align}
Performing the same method used to solve Eq.~(\ref{glEOM}), we obtain the solution of Eq.~(\ref{mcEOM});
\begin{align} \nonumber
F^X=&-\bar{f}\left[1+\frac{1}{|f|^2}\left(\frac{\bar{G}^2}{2f}\right)\left\{\partial^2\left(\frac{G^2}{2\bar{f}}\right)+\frac{1}{2}\bar{f}''\bar{\chi}^2+\bar{f}'g'\right\}\right.\\ \nonumber
&+\frac{G^2\bar{G}^2}{4|f|^6}\left\{|f'f|^2-2|\bar{f}''\bar{\chi}^2/2+\bar{f}'g'|^2-\left(\partial^2\left(\frac{G^2}{2\bar{f}}\right)(f''\chi^2/2+f'\bar{g}')+h.c.\right)\right\}\\ 
&\left. -\frac{3}{16|f|^8}G^2\bar{G}^2\partial^2G^2\partial^2\bar{G}^2\right].
\end{align}
Substituting the solution into Eq.~(\ref{mcVAac2}), we obtain the on-shell Lagrangian in Eq.~(\ref{mcVAac});
\begin{align}
{\mathcal L}=&-(|f|^2+|g'|^2)-\partial_{a}\phi\partial^{a}\bar{\phi}+i\partial_a\bar{\chi}\bar{\sigma}^a \chi+i\partial_a\bar{G}\bar{\sigma}^a G\nonumber\\
&-\left(f'G\chi+\frac{1}{2}g''\chi^2-\frac{f'\bar{g}'}{2\bar{f}}G^2+h.c.\right) \nonumber\\
&+\frac{\bar{G}^2}{2f}\partial^2\left(\frac{G^2}{2\bar{f}}\right)-\frac{|f'|^2}{4|f|^4}(|f|^2+|g'|^2)G^2\bar{G}^2+\left(\frac{f''}{4\bar{f}}G^2\chi^2+h.c.\right) \nonumber\\
&-\left(\frac{f'\bar{f}''\bar{g}'}{8|f|^4}\bar{\chi}^2G^2\bar{G}^2+\frac{ff'\bar{g}'}{8|f|^6}G^2\bar{G}^2\partial^2G^2+h.c.\right)\nonumber\\
&-\frac{1}{16|f|^6}G^2\bar{G}^2\partial^2G^2\partial^2\bar{G}^2-\frac{|f''|^2}{16|f|^4}G^2\bar{G}^2\chi^2\bar{\chi}^2-\left(\frac{ff''}{16|f|^6}G^2\bar{G}^2\chi^2\partial^2G^2+h.c.\right).\label{glmcVA}
\end{align}
The Lagrangian~(\ref{glmcVA}) has not only the higher order terms of $G^\alpha$ as in the previous case but also the higher order interactions between $G^\alpha$ and the matter fermion $\chi$. Such higher order corrections are suppressed by $\sqrt{|\langle f\rangle|}$, which would be the order parameter of the SUSY breaking. If SUSY breaking scale is sufficiently high compared to the energy scale of collider experiments, the effects of those interactions are not important, as discussed in Ref.~\cite{Antoniadis:2010hs}. However, in the early universe, such as the reheating era, the energy scale of the momenta is much higher than $\sqrt{|\langle f\rangle|}$. In such a case, the higher order terms shown in Eq.~(\ref{glmcVA}) may become important.

\section{VA multiplet action in SUGRA}
\subsection{Case without matter multiplets}
In the previous section, we have derived the action of the VA multiplet in global SUSY. In the following, we discuss the one coupled to SUGRA. Even in such a case, we can derive the on-shell action through the same procedure we did in the previous section.

To construct the action, we use the conformal SUGRA formulation~\cite{Kaku:1978nz,Kaku:1978ea,Townsend:1979ki,Kugo:1982cu},\footnote{For review, see Ref.~\cite{supergravity}.} with which we can avoid tedious field redefinition procedure associated with Poincar\'e SUGRA. In this section, we use the notation in the Ref.~\cite{supergravity}. In conformal SUGRA, the constraint on the VA multiplet can be written as in the global SUSY case:
\begin{align}
\hat{X}^2=0,\label{sgcon}
\end{align}
where $\hat{X}$ is the chiral VA multiplet whose Weyl and chiral weights, denoted by $w$, and $n$ respectively, are $(w,n)=(0,0)$.\footnote{The Weyl and the chiral weights are the charges under the dilatation and U$(1)_T$, which are parts of the superconformal symmetry.} We can solve Eq.~(\ref{sgcon}) and obtain
\begin{align}
X=\frac{\bar{G}P_LG}{2F^X},
\end{align}
where $X$, $P_LG$, and $F^X$ are the scalar, the fermion, the auxiliary components of $\hat{X}$ respectively, and $P_L$ is a chirality projection operator. The action of the chiral multiplet is generically given by
\begin{align}
S=\left[-3S_0\bar{S}_0e^{-K/3}\right]_D+[S_0^3W]_F,\label{SGaction}
\end{align}
where $S_0$ denotes the chiral compensator with $(w,n)=(1,1)$, $[\cdots]_{D,F}$ denote the D- and F-term density formulae respectively. $K$ and $W$ are K\"ahler and superpotential respectively. Here we choose the following $K$ and $W$,
\begin{align}
K=&|X|^2,\\
W=&fX+W_0,
\end{align}
where $f$ and $W_0$ are complex constants. To fix unphysical parts of the superconformal symmetry, we put the following conventional gauge conditions~\cite{Kugo:1982mr} on the action~(\ref{SGaction}),
\begin{align}
S_0=\bar{S}_0=e^{K/6},\quad P_L\chi^0-\frac{1}{3}e^{K/6}\bar{X}P_LG=0,\quad b_\mu=0 \label{SCgauge}
\end{align}
where we use the Planck unit convention ($M_{pl}\sim2.4\times 10^{18}{\rm GeV}=1$), the first condition corresponds to the dilatation and U$(1)_T$ ones, the second corresponds to the S-SUSY one, $P_L\chi^0$ denotes the fermionic component of $S_0$, $b_\mu$ is the gauge field of dilatation, and the last condition corresponds to the special conformal boost one. Under these gauge conditions, the action (\ref{SGaction}) contains the auxiliary fields $A_\mu$ and $F^0$ which are the U$(1)_T$ gauge field and the auxiliary component of $S_0$ respectively. We can eliminate them by E.O.M of them, and then we obtain the following action: 
\begin{align}
\mathcal{L}=&\frac{1}{2}(R-\bar{\psi}_\mu\mathcal{R}^\mu+\mathcal{L}_{SGT})-\partial_\mu X\partial^\mu \bar{X}-\frac{1}{2}(\bar{G}P_L\slashed{\partial}G+\bar{G}P_R\slashed{\partial}G)-\frac{1}{4}(\bar{G}P_L\slashed{\partial}XG\bar{X}+{\rm h.c.})\nonumber\\
&-\frac{1}{8}(\bar{G}P_LG)(\bar{G}P_RG)-\frac{e^{K/2}}{2}\left\{\left(2f\bar{X}+\bar{X}^2W\right)+{\rm h.c.}\right\}\nonumber\\
&+3e^K|W|^2+|F^X|^2+\left(e^{K/2}D_XWF^X+{\rm h.c.} \right)\nonumber\\
&+\frac{1}{\sqrt{2}}\left\{\bar{\psi}_\mu\slashed{\partial}\bar{X}\gamma^\mu P_LG+{\rm h.c.}\right\}-\frac{1}{2}(\bar{\psi}_\mu P_LG)(\bar{\psi}^\mu P_RG)\nonumber\\
&+\frac{i}{16}\epsilon^{abcd}(\bar{\psi}_a\gamma_b\psi_c)\bar{G}P_R\gamma_dG+\frac{i}{8}\epsilon^{abcd}(\bar{\psi}_a\gamma_b\psi_c)(\bar{X}\partial_dX-X\partial_d\bar{X})\nonumber\\
&\left\{\frac{1}{\sqrt{2}}e^{K/2}D_XW\bar{\psi}_\mu\gamma^\mu P_LG+\frac{1}{2}e^{K/2}W\bar{\psi}_\mu P_R\gamma^{\mu\nu}\psi_\nu+{\rm h.c.}\right\},\label{offSG}
\end{align}
where
\begin{align}
\mathcal{R}^\mu&=\gamma^{\mu\nu\rho}\left(\partial_\nu+\frac{1}{4}\omega_{\nu}^{ab}\gamma_{ab}\right)\psi_\rho,\\
\mathcal{L}_{SGT}&=\frac{1}{16}\left[-(\bar{\psi}_\mu\gamma_\nu\psi_\rho)(\bar{\psi}^\mu\gamma^\nu\psi^\rho)-2(\bar{\psi}_\mu\gamma_\nu\psi_\rho)(\bar{\psi}^\mu\gamma^\rho\psi^\nu)+4(\bar{\psi}_\mu\gamma^\mu\psi_\nu)(\bar{\psi}_\rho\gamma^\rho\psi^\nu)\right]
\end{align}
Taking into account the fact that $X=\bar{G}P_LG/(2F^X)$, we obtain the following E.O.M by the variation of $\bar{F}^{\bar{X}}$,
\begin{align}
&F^X+\bar{f}+\bar{W}_0X+\frac{3}{2}\bar{f}|X|^2-\Biggl[\Box X+3W_0\bar{f}+3|f|^2X-f\bar{G}P_LG+\left(W_0+\frac{3}{2}fX\right)F^X\nonumber\\
&+\frac{3}{2}\bar{f}X\bar{F}^{\bar{X}}-\frac{1}{\sqrt{2}}\nabla_\nu(\bar{\psi}_\mu\gamma^\nu\gamma^\mu P_LG)+\frac{i}{8}\epsilon^{\mu\nu\rho\sigma}\bar{\psi}_\mu\gamma_\nu\psi_\rho\partial_\sigma X+\frac{i}{8}\nabla_\sigma(\epsilon^{\mu\nu\rho\sigma}\bar{\psi}_\mu\gamma_\nu\psi_\sigma X)\nonumber\\
&+\frac{1}{\sqrt{2}}W_0\bar{\psi}^\mu\gamma_\mu P_LG+\frac{1}{2}W_0X\bar{\psi}_\mu P_R\gamma^{\mu\nu}\psi_\nu+\frac{1}{2}(\bar{f}+\bar{W}_0X)\bar{\psi}_\mu P_L\gamma^{\mu\nu}\psi_\nu\Biggr]\frac{\bar{X}}{\bar{F}^X}=0.\label{SGEOM}
\end{align}
As in the case of Sec.~\ref{global}, we can substitute the ansatz~(\ref{ansatz}) to the E.O.M, and obtain the following set of equations from the each coefficient of the series of $\bar{G}P_LG$ and $\bar{G}P_RG$,
\begin{align}
A=&-\bar{f},\\
B=&-\frac{\bar{W}_0}{2A},\\
C=&\frac{1}{2\bar{A}^2}\Biggl[\Box \left(\frac{\bar{G}P_LG}{2A}\right)+3W_0\bar{f}+W_0A-\frac{1}{\sqrt{2}}\nabla_\nu(\bar{\psi}_\mu\gamma^\nu\gamma^\mu P_LG)\nonumber\\
&+\frac{i}{4}\epsilon^{\mu\nu\rho\sigma}\bar{\psi}_\mu\gamma_\nu\psi_\rho\nabla_\sigma\left(\frac{\bar{G}P_LG}{2A}\right)+\frac{1}{\sqrt{2}}W_0\bar{\psi}_\mu\gamma^\mu P_LG+\frac{1}{2}\bar{f}\bar{\psi}_\mu P_L\gamma^{\mu\nu}\psi_\nu\Biggr],\\
D=&\frac{CW_0}{2A^2}-\frac{C\nabla_\mu\bar{G}P_R\nabla^\mu G}{4|A|^4}+\frac{BW_0}{2\bar{A}^2}\nonumber\\
&-\frac{\bar{C}^2}{\bar{A}^3}\left[\frac{\nabla_\mu \bar{G}P_L\nabla^\mu G}{2A}+3W_0\bar{f}+W_0\bar{f}+W_0A-\frac{1}{\sqrt{2}}\bar{\psi}_\mu\gamma_\nu\gamma^\mu\nabla^\nu P_LG+\frac{1}{2}\bar{f}\bar{\psi}_\mu P_L\gamma^{\mu\nu}\psi_\nu\right]\nonumber\\
&-\frac{1}{4|A|^2}\Biggl[-\frac{3|f|^2}{\bar{A}}+\frac{fA}{2\bar{A}}-\frac{i}{8\bar{A}}\epsilon^{\mu\nu\rho\sigma}\nabla_\sigma(\bar{\psi}_\mu\gamma_\nu\psi_\rho)-\frac{W_0}{2\bar{A}}\bar{\psi}_\mu P_R\gamma^{\mu\nu}\psi_\nu-\frac{\bar{W}_0}{2\bar{A}}\bar{\psi}_\mu P_L\gamma^{\mu\nu}\psi_\nu\Biggr].
\end{align}
These equations uniquely determine the value of $A$, $B$, $C$, and $D$. Thus, we can obtain the solution of the E.O.M~(\ref{SGEOM}). The explicit solution for $F^X$ is complicated, and so we omit it here.

By substituting the on-shell expression of $F^X$ into Eq.~(\ref{offSG}), we obtain the on-shell Lagrangian as\footnote{We would like to Magnus Tournoy for pointing out the last term missing in our previous version.}
\begin{align}
\mathcal{L}=&\frac{1}{2}(R-\bar{\psi}_\mu\mathcal{R}^\mu+\mathcal{L}_{SGT})-\frac{1}{2}(\bar{G}P_L\slashed{\nabla}G+\bar{G}P_R\slashed{\nabla}G)-|f|^2+3|W_0|^2-\frac{1}{8}(\bar{G}P_LG)(\bar{G}P_RG)\nonumber\\
&+\left[-\frac{f\bar{W}_0}{\bar{f}}\bar{G}P_LG+\frac{1}{\sqrt{2}}f\bar{\psi}_\mu\gamma^\mu P_LG+\frac{W_0}{2}\bar{\psi}_\mu P_R\gamma^{\mu\nu}\psi_\nu+{\rm h.c.}\right]\nonumber\\
&-\frac{1}{4|f|^2}\nabla_\mu(\bar{G}P_LG)\nabla^\mu(\bar{G}P_RG)-\frac{1}{2}(\bar{\psi}_\mu P_LG)(\bar{\psi}^\mu P_RG)+\frac{i}{16}\epsilon^{\mu\nu\rho\sigma}(\bar{\psi}_\mu\gamma_\nu \psi_\rho)(\bar{G}P_R\gamma_\sigma G)\nonumber\\
&-\left(\frac{W_0}{2\sqrt{2}f}\bar{G}P_RG\bar{\psi}_\mu\gamma^\mu P_LG+\frac{f}{4\bar{f}}\bar{G}P_LG\bar{\psi}_\mu P_R\gamma^{\mu\nu}\psi_\nu+{\rm h.c.}\right)\nonumber\\
&-\frac{1}{2\sqrt{2}}\left[\frac{1}{f}\bar{\psi}_\mu\slashed{\nabla}(\bar{G}P_RG)\gamma^\mu P_LG
+{\rm h.c.}\right]\nonumber\\
&+\frac{i}{32|f|^2}\epsilon^{\mu\nu\rho\sigma}(\bar{\psi}_\mu\gamma_\nu\psi_\rho)\left(\bar{G}P_RG\nabla_\sigma(\bar{G}P_LG)-\bar{G}P_LG\nabla_\sigma(\bar{G}P_RG)\right)\nonumber\\
&+\frac{(\bar{G}P_LG)(\bar{G}P_RG)}{4|f|^2}\left[-4|f|^2|\hat C|^2+4|f|^2+2|W_0|^2+\left\{\frac{W_0}{4}\bar{\psi}_\mu P_R\gamma^{\mu\nu}\psi_\nu+\text{h.c.}\right\}\right]\nonumber\\\label{VASUGRA}
\end{align}
where $\hat{C}$ on the fifth line is defined as
\begin{align}
\hat{C}\equiv\frac{1}{2f^2}\left[2W_0\bar{f}-\frac{\nabla_\mu\bar{G}P_L\nabla^\mu G}{2\bar{f}}-\frac{1}{\sqrt{2}}\bar{\psi}_\mu\gamma^\nu\gamma^\mu\nabla_\nu P_LG+\frac{\bar{f}}{2}\bar{\psi}_\mu P_L\gamma^{\mu\nu}\psi_\nu\right].
\end{align}

The fermion $P_LG$ is the Goldstino in this system, and therefore, we take the unitary gauge $P_LG=0$. Under the gauge condition, the Lagrangian~(\ref{VASUGRA}) takes a very simple form given by
\begin{align}
\mathcal{L}=&\frac{1}{2}(R-\bar{\psi}_\mu\mathcal{R}^\mu+\mathcal{L}_{SGT})-|f|^2+3|W_0|^2+\left[\frac{W_0}{2}\bar{\psi}_\mu P_R\gamma^{\mu\nu}\psi_\nu+{\rm h.c.}\right].\label{unitary}
\end{align}
The Lagrangian~(\ref{unitary}) corresponds to the one in Ref.~\cite{Lindstrom:1979kq} if we set $W_0=0$. This Lagrangian describes the pure supergravity system containing a graviton, a massive gravitino with $|m_{3/2}|=|W_0|$, and a cosmological constant $\Lambda=-|f|^2+3|W_0|^2$.
\subsection{Case with a matter multiplet}
Let us extend the result shown in the previous subsection to the matter coupled one. Although, in principle, we can construct a system with multiple chiral matter multiplets and gauge multiplets, we discuss a simple case where there is a single chiral matter multiplet $\Phi$. We assume the following K\"ahler and superpotential,
\begin{align}
K=&\hat{K}(\Phi,\bar{\Phi})+|X|^2,\\
W=&f(\Phi)X+g(\Phi),
\end{align}
where $\hat{K}(\Phi,\bar{\Phi})$ is a real function of $\Phi$ and $\bar{\Phi}$, $f(\Phi)$ and $g(\Phi)$ are holomorphic functions of $\Phi$. We can obtain the off-shell Poincar\'e SUGRA action by substituting $K$ and $W$ into Eq.~(\ref{SGaction}) with the following superconformal gauge conditions, instead of the ones in Eq.~(\ref{SCgauge}),
\begin{align}
S_0=\bar{S}_0=e^{K/6},\quad P_L\chi^0-\frac{1}{3}e^{K/6}(\bar{X}P_LG+\hat{K}_\Phi P_L\chi)=0, \quad b_\mu=0,
\end{align}
where $\hat{K}_\Phi$ denotes the derivative of $\hat{K}$ with respect to $\Phi$.

The off-shell Lagrangian is 
\begin{align}
\mathcal{L}=&\frac{1}{2}(R-\bar{\psi}_\mu\mathcal{R}^\mu+\mathcal{L}_{SGT})-K_{I\bar{J}}\nabla_\mu z^I\nabla^\mu \bar{z}^{\bar{J}}-\frac{1}{2}K_{I\bar{J}}(\bar{\chi}^IP_L\slashed{\nabla}\chi^{\bar{J}}+\bar{\chi}^{\bar{J}}P_R\slashed{\nabla}\chi^I)\nonumber\\
&\left\{\frac{1}{2}\bar{\chi}^JP_L\slashed{\nabla}z^I\chi^{\bar{K}}(K_{IJ\bar{K}}-\frac{1}{2}K_{I}K_{J\bar{K}})+{\rm h.c.}\right\}+\frac{1}{4}(\bar{\chi}^IP_L\chi^J)(\bar{\chi}^{\bar{K}}P_R\chi^{\bar{L}})(K_{IJ\bar{K}\bar{L}}-\frac{1}{2}K_{I\bar{K}}K_{J\bar{L}})\nonumber\\
&+\left\{-\frac{1}{2}e^{\frac{K}{2}}(W_{IJ}+K_IW_J+K_JW_I+K_{IJ}W+K_IK_JW)\bar{\chi}^IP_L\chi^J+{\rm h.c.}\right\}\nonumber\\
&+\left\{\frac{1}{\sqrt{2}}K_{I\bar{J}}\bar{\psi}_\mu\slashed{\nabla}\bar{z}^{\bar{J}}\gamma^\mu P_L\chi^I+{\rm h.c.}\right\}-\frac{1}{2}K_{I\bar{J}}(\bar{\psi}_\mu P_L\chi^I)(\bar{\psi}^\mu P_R\chi^{\bar{J}})\nonumber\\
&\frac{i}{16}\epsilon^{\mu\nu\rho\sigma} K_{I\bar{J}}(\bar{\psi}_\mu\gamma_\nu\psi_\rho)\bar{\chi}^{\bar{J}}P_R\gamma_{\sigma}\chi^I+\frac{i}{8}\epsilon^{\mu\nu\rho\sigma}(\bar{\psi}_\mu\gamma_{\nu}\psi_\rho)(K_I\nabla_{\sigma}z^I-K_{\bar{J}}\bar{z}^{\bar{J}})\nonumber\\
&\left\{\frac{1}{2}e^{\frac{K}{2}}D_IW\bar{\psi}_\mu\gamma^{\mu}P_L\chi^{I}+\frac{1}{2}e^{\frac{K}{2}}W\bar{\psi}_\mu P_R\gamma^{\mu\nu}\psi_\nu+{\rm h.c.}\right\}\nonumber\\
&+3e^{K}|W|^2+K_{I\bar{J}}F^I\bar{F}^{\bar{J}}-\frac{1}{2}(\bar{F}^{\bar{K}}\bar{\chi}^IP_L\chi^JK_{IJ\bar{K}}+{\rm h.c.})+(e^{\frac{K}{2}}D_IWF^I),
\end{align}
where $I,J\cdots=\Phi,X$, $z^\Phi=\phi$, $z^X=X$, $\chi^\Phi=\psi$, and $\chi^X=G$ and $D_IW\equiv W_I+K_IW$.\\
As in the same way performed in Sec.~2 and 3.1, we can solve the E.O.M of the auxiliary field $F^X$ straightforwardly. Therefore, we just show the resultant on-shell action: 
\begin{align}
\mathcal{L}=&\frac{1}{2}(R-\bar{\psi}_\mu\mathcal{R}^\mu+\mathcal{L}_{SGT})-K_{\Phi\bar\Phi}\partial_\mu\Phi\partial^\mu\bar\Phi-\frac{1}{2}K_{\Phi\bar\Phi}(\bar{\chi}P_L\slashed{\nabla}\chi+\bar{\chi}P_R\slashed{\nabla}\chi)\nonumber\\
&-\frac{1}{2}(\bar{G}P_L\slashed{\nabla}G+\bar{G}P_R\slashed{\nabla}G)-|\hat f|^2+3|\hat g|^2-K^{\Phi\bar\Phi}|\tilde{m}_{G\chi}|^2\nonumber\\
&\left\{\frac{1}{2}\bar{\chi}P_L\slashed{\nabla}\Phi\chi(K_{\Phi\Phi\bar{\Phi}}-\frac{1}{2}K_\Phi K_{\Phi\bar{\Phi}})-\frac{1}{4}K_\Phi\bar GP_L\slashed{\partial}\Phi G+{\rm h.c.}\right\}\nonumber\\
&+\frac{1}{4}(\bar{\chi}P_L\chi)(\bar{\chi}P_R\chi)(K_{\Phi\Phi\bar{\Phi}\bar{\Phi}}-\frac{1}{2}K_{\Phi\bar{\Phi}}K_{\Phi\bar{\Phi}}-K_{\Phi\Phi\bar\Phi}K^{\Phi\bar\Phi}K_{\Phi\bar\Phi\bar\Phi})\nonumber\\
&-\frac{1}{4}K_{\Phi\bar\Phi}(\bar{G}P_L\chi)(\bar{G}P_R\chi)-\frac{1}{8}(\bar{G}P_LG)(\bar{G}P_RG)\nonumber\\
&-\frac{1}{2}(m_{\chi\chi}\bar{\chi}P_L\chi+m_{G\chi}\bar{G}P_L\chi+{\rm h.c.})+\frac{1}{\sqrt 2}K_{\Phi\bar\Phi}(\bar\psi_\mu\slashed\partial\bar\Phi\gamma^\mu P_L\chi+{\rm h.c.})\nonumber\\
&-\frac{1}{2}K_{\Phi\bar\Phi}(\bar{\psi}_\mu P_L\chi)(\bar{\psi}^\mu P_R\chi)-\frac{1}{2}(\bar{\psi}_\mu P_LG)(\bar{\psi}^\mu P_RG)\nonumber\\
&+\frac{i}{16}\epsilon^{\mu\nu\rho\sigma}K_{\Phi\bar\Phi}\bar\psi_\mu \gamma_\nu \psi_\rho \bar\chi P_R\gamma_\sigma \chi+\frac{i}{16}\epsilon^{\mu\nu\rho\sigma}\bar\psi_\mu \gamma_\nu \psi_\rho \bar G P_R\gamma_\sigma G+\frac{i}{8}\epsilon^{\mu\nu\rho\sigma}\bar\psi_\mu \gamma_\nu \psi_\rho (K_{\Phi}\partial_\sigma \Phi-K_{\bar\Phi}\partial_\sigma \bar\Phi)\nonumber\\
&+\frac{i}{32|f|^2}\epsilon^{\mu\nu\rho\sigma}(\bar{\psi}_\mu\gamma_\nu\psi_\rho)\left(\bar{G}P_RG\nabla_\sigma(\bar{G}P_LG)-\bar{G}P_LG\nabla_\sigma(\bar{G}P_RG)\right)\nonumber\\
&+\left\{\frac{1}{\sqrt{2}}\tilde{m}_{G\chi}\bar{\psi}\cdot\gamma P_L\chi+\frac{1}{\sqrt{2}}\hat f\bar{\psi}\cdot\gamma P_LG+\frac{1}{2}\hat g\bar\psi_\mu P_R\gamma^{\mu\nu}\psi_\nu+{\rm h.c.}\right\}\nonumber\\
&-\frac{1}{4}\nabla_\mu\left(\frac{\bar{G}P_LG}{\overline{\hat{f}}}\right)\nabla^\mu\left(\frac{\bar{G}P_RG}{\hat{f}}\right)-\frac{1}{16}\left\{\frac{1}{\hat f}\bar{G}P_RG\bar\chi R_L\gamma^\mu\chi\nabla_\mu\left(\frac{\bar{G}P_LG}{\overline{\hat{f}}}\right)+{\rm h.c.}\right\}\nonumber\\
&\left(-\frac{\hat f\hat g}{\overline{\hat{f}}}\bar{G}P_LG+\frac{1}{4\overline{\hat{f}}}\tilde{m}_{\chi\chi}\bar{\chi}P_L\chi \bar{G}P_LG+\frac{1}{4{\hat{f}}}\tilde{m}_{\chi G}\bar{G}P_L\chi \bar{G}P_RG+{\rm h.c.}\right)\nonumber\\
&-\frac{1}{2\sqrt 2}\left\{m_{G\chi}\frac{\bar{G}P_LG}{\overline{\hat{f}}}\bar\psi\cdot\gamma P_L \chi+\hat g\frac{\bar{G}P_RG}{\hat{f}}\bar\psi\cdot\gamma P_L G+{\rm h.c.}\right\}\nonumber\\
&-\left(\frac{\hat f}{4\overline{\hat{f}}}\bar{G}P_LG\bar\psi_\mu P_R \gamma^{\mu\nu}\psi_\nu+{\rm h.c.}\right)+\frac{K^{\Phi\bar\Phi}}{2}\left\{\tilde{m}_{G\chi}\overline{m_{G\chi}}\frac{\bar{G}P_RG}{\hat{f}}+{\rm h.c.}\right\}\nonumber\\
&\frac{(\bar{G}P_LG)(\bar{G}P_RG)}{4|\hat f|^2}\left[-4|\hat f|^2|\mathcal{\hat C}|^2+2|\hat g|^2+4|\hat f|^2-K^{\Phi\bar\Phi}| {m}_{G\chi}|^2-\frac{1}{4}(m_{\chi\chi}\bar{\chi}P_L\chi+{\rm h.c.})\right. \nonumber\\
&\hspace{3.3cm}\left.+\left\{\frac{1}{2\sqrt 2}\tilde{m}_{G\chi}\bar{\psi}\cdot\gamma P_L\chi+\frac{1}{4}\hat g\bar\psi_\mu P_R\gamma^{\mu\nu}\psi_\nu+{\rm h.c.}\right\}\right],\label{mVASUGRA}
\end{align}
where
\begin{align}
&\hat f=e^{\hat K/2}f,\\
&\hat g=e^{\hat K/2}g,\\
&m_{\chi\chi}=e^{\hat K/2}[g''+2K_\Phi g'+K_{\Phi\Phi}g+K_\Phi K_\Phi g-\Gamma^\Phi_{\Phi\Phi}(g'+K_\Phi g)],\\
&m_{G\chi}=e^{\hat K/2}(f'+K_\Phi f),\\
&\tilde{m}_{\chi\chi}=e^{\hat K/2}[\bar f''+2K_{\bar\Phi}\bar f'+K_{\bar\Phi\bar\Phi}\bar f+K_{\bar\Phi}K_{\bar\Phi}\bar f-\Gamma^{\bar\Phi}_{\bar\Phi\bar\Phi}(\bar f'+K_{\bar\Phi} f)],\\
&\tilde{m}_{G\chi}=e^{\hat K/2}(g'+K_\Phi g),\\
&\hspace{0.6cm}\mathcal{\hat C}=\frac{1}{2{\hat f}^2}\Biggl[2\hat g\overline{\hat f}-\frac{1}{2}\tilde{m}_{\chi\chi}\bar\chi P_R \chi-\frac{1}{2\sqrt 2}\bar\psi_\nu\gamma^\mu\gamma^\nu(\nabla_\mu P_LG)+\frac{1}{2\sqrt 2}\overline{m_{G\chi}}\bar\psi\cdot\gamma P_R\chi\nonumber\\
&\hspace{2cm}+\frac{1}{2}\overline{\hat f}\bar\psi_\mu P_L\gamma^{\mu\nu}\psi_\nu-\frac{\nabla_\mu\bar{G}P_L\nabla^\mu G}{2\overline{\hat f}}\Biggr].
\end{align}

\section{Summary}
In this work, we have constructed the action of the VA multiplet coupled to matter multiplet in global SUSY and SUGRA. The scalar component of the VA multiplet is related to its fermionic and auxiliary components as shown in Eq.~(\ref{const}) due to the nilpotent condition $\hat{X}^2=0$. Such a constraint makes the E.O.M of the auxiliary field complicated, however, nonlinear terms in E.O.M are coupled to the fermion bilinear $\bar{G}P_LG$ and (or) $\bar{G}P_R G$. Because of such a special structure of E.O.M, we can solve it in a systematic way, which we have performed in Sec.~2 and 3.

As shown in Eqs.~(\ref{glmcVA}) and (\ref{mVASUGRA}), in cases with a matter multiplet, the higher order interactions between the VA fermion and the matter fermion appear, which are suppressed by not the Planck mass but $\sqrt{|\langle f\rangle|}$. As we mentioned, such couplings may become important for high scale physics, such as the reheating after inflation and the gravitino production during it. We expect that our construction is useful for a study of the phenomenological and cosmological consequences of the VA multiplet. For example, the action coupled to SUGRA enable us to discuss the perturbative and non-perturbative productions of gravitino as discussed in Refs.~\cite{Kallosh:1999jj,Kallosh:2000ve}

The Dirac-Born-Infeld type action in 4 dimensional ${\cal N}=1$ SUGRA is also described by the system with the VA and gauge multiplets~\cite{Abe:2015nxa}. Our method is also applicable to constructing the component expression of such an action, which contain the higher order terms of gaugino. We will study such a system elsewhere.

\section*{\bf Note added}
While we were completing this work, the paper~\cite{Bergshoeff:2015tra} by Bergshoeff et al. appeared. They also discussed the model in Sec.3.1 of this paper. In Ref.~\cite{Bergshoeff:2015tra}, they used the superconformal gauge condition which is different from ours~(\ref{SCgauge}). With our choice of the gauge conditions, the coupling between Ricci scalar and the VA fermion is absent, while the Lagrangian shown in Ref.~\cite{Bergshoeff:2015tra} contains such a non-minimal coupling, therefore the form of the Lagrangian seems different.

\subsection*{Acknowledgements}
We are grateful to Magnus Tournoy for pointing out a typo in the previous version of this paper. Y.Y. are supported in part by Research Fellowships for Young Scientists (No.26-4236), 
which are from Japan Society for the Promotion of Science.  


\begin{thebibliography}{99}
\bibitem{Volkov:1973ix} 
  D.~V.~Volkov and V.~P.~Akulov,
  Phys.\ Lett.\ B {\bf 46}, 109 (1973).
\bibitem{Ivanov:1978mx} 
  E.~A.~Ivanov and A.~A.~Kapustnikov,
  J.\ Phys.\ A {\bf 11}, 2375 (1978).
\bibitem{Rocek:1978nb} 
  M.~Rocek,
  Phys.\ Rev.\ Lett.\  {\bf 41}, 451 (1978).
\bibitem{Antoniadis:2014oya} 
  I.~Antoniadis, E.~Dudas, S.~Ferrara and A.~Sagnotti,
  Phys.\ Lett.\ B {\bf 733}, 32 (2014)
  [arXiv:1403.3269 [hep-th]].
\bibitem{Ferrara:2014kva} 
  S.~Ferrara, R.~Kallosh and A.~Linde,
  JHEP {\bf 1410}, 143 (2014)
  [arXiv:1408.4096 [hep-th]].
\bibitem{Kallosh:2014via} 
  R.~Kallosh and A.~Linde,
  JCAP {\bf 1501}, no. 01, 025 (2015)
  [arXiv:1408.5950 [hep-th]].
\bibitem{Aoki:2014pna} 
  S.~Aoki and Y.~Yamada,
  Phys.\ Rev.\ D {\bf 90}, no. 12, 127701 (2014)
  [arXiv:1409.4183 [hep-th]].
\bibitem{Dall'Agata:2014oka} 
  G.~Dall'Agata and F.~Zwirner,
  JHEP {\bf 1412}, 172 (2014)
  [arXiv:1411.2605 [hep-th]].
\bibitem{Kallosh:2014hxa} 
  R.~Kallosh, A.~Linde and M.~Scalisi,
  JHEP {\bf 1503}, 111 (2015)
  [arXiv:1411.5671 [hep-th]].
\bibitem{Kallosh:2015lwa} 
  R.~Kallosh and A.~Linde,
  Phys.\ Rev.\ D {\bf 91}, 083528 (2015)
  [arXiv:1502.07733 [astro-ph.CO]].
\bibitem{Scalisi:2015qga} 
  M.~Scalisi,
  arXiv:1506.01368 [hep-th].
\bibitem{Carrasco:2015pla} 
  J.~J.~M.~Carrasco, R.~Kallosh and A.~Linde,
  arXiv:1506.01708 [hep-th].
\bibitem{Choi:2005ge} 
  K.~Choi, A.~Falkowski, H.~P.~Nilles and M.~Olechowski,
  Nucl.\ Phys.\ B {\bf 718}, 113 (2005)
  [hep-th/0503216].
\bibitem{Kallosh:2014wsa} 
  R.~Kallosh and T.~Wrase,
  JHEP {\bf 1412}, 117 (2014)
  [arXiv:1411.1121 [hep-th]].
\bibitem{Bergshoeff:2015jxa} 
  E.~A.~Bergshoeff, K.~Dasgupta, R.~Kallosh, A.~Van Proeyen and T.~Wrase,
  JHEP {\bf 1505}, 058 (2015)
  [arXiv:1502.07627 [hep-th]].
\bibitem{Kallosh:2015nia} 
  R.~Kallosh, F.~Quevedo and A.~M.~Uranga,
  arXiv:1507.07556 [hep-th].
\bibitem{Kachru:2003aw} 
  S.~Kachru, R.~Kallosh, A.~D.~Linde and S.~P.~Trivedi,
  Phys.\ Rev.\ D {\bf 68}, 046005 (2003)
  [hep-th/0301240].
\bibitem{Balasubramanian:2005zx} 
  V.~Balasubramanian, P.~Berglund, J.~P.~Conlon and F.~Quevedo,
  JHEP {\bf 0503}, 007 (2005)
  [hep-th/0502058].
\bibitem{Antoniadis:2010hs} 
  I.~Antoniadis, E.~Dudas, D.~M.~Ghilencea and P.~Tziveloglou,
  Nucl.\ Phys.\ B {\bf 841}, 157 (2010)
  [arXiv:1006.1662 [hep-ph]].
\bibitem{Dudas:2011kt} 
  E.~Dudas, G.~von Gersdorff, D.~M.~Ghilencea, S.~Lavignac and J.~Parmentier,
  Nucl.\ Phys.\ B {\bf 855}, 570 (2012)
  [arXiv:1106.5792 [hep-th]].
\bibitem{Antoniadis:2011xi} 
  I.~Antoniadis, E.~Dudas and D.~M.~Ghilencea,
  Nucl.\ Phys.\ B {\bf 857}, 65 (2012)
  [arXiv:1110.5939 [hep-th]].
\bibitem{Antoniadis:2012zz} 
  I.~Antoniadis, E.~Dudas, D.~M.~Ghilencea and P.~Tziveloglou,
  Theor.\ Math.\ Phys.\  {\bf 170}, 26 (2012)
  [Teor.\ Mat.\ Fiz.\  {\bf 170}, 34 (2012)].
\bibitem{Farakos:2012fm} 
  F.~Farakos and A.~Kehagias,
  Phys.\ Lett.\ B {\bf 719}, 95 (2013)
  [arXiv:1210.4941 [hep-ph]].
\bibitem{Komargodski:2009rz} 
  Z.~Komargodski and N.~Seiberg,
  JHEP {\bf 0909}, 066 (2009)
  [arXiv:0907.2441 [hep-th]].
\bibitem{Kuzenko:2011tj} 
  S.~M.~Kuzenko and S.~J.~Tyler,
  JHEP {\bf 1105}, 055 (2011)
  [arXiv:1102.3043 [hep-th]].
\bibitem{Casalbuoni:1988xh} 
  R.~Casalbuoni, S.~De Curtis, D.~Dominici, F.~Feruglio and R.~Gatto,
  Phys.\ Lett.\ B {\bf 220}, 569 (1989).
\bibitem{Farakos:2013ih} 
  F.~Farakos and A.~Kehagias,
  Phys.\ Lett.\ B {\bf 724}, 322 (2013)
  [arXiv:1302.0866 [hep-th]].
\bibitem{Kuzenko:2011ti} 
  S.~M.~Kuzenko and S.~J.~Tyler,
  JHEP {\bf 1104}, 057 (2011)
  [arXiv:1102.3042 [hep-th]].
\bibitem{Farakos:2015vba} 
  F.~Farakos, O.~Hul$\rm \acute{\i}$k, P.~Ko$\rm\check{c}$$\acute{\i}$ and R.~von Unge,
  arXiv:1507.01885 [hep-th].
\bibitem{Kuzenko:2015uca} 
  S.~M.~Kuzenko and S.~J.~Tyler,
  arXiv:1507.04593 [hep-th].
\bibitem{Lindstrom:1979kq} 
  U.~Lindstrom and M.~Rocek,
  Phys.\ Rev.\ D {\bf 19}, 2300 (1979).
\bibitem{Samuel:1982uh} 
  S.~Samuel and J.~Wess,
  Nucl.\ Phys.\ B {\bf 221}, 153 (1983).
\bibitem{Samuel:1983ui} 
  S.~Samuel and J.~Wess,
  Nucl.\ Phys.\ B {\bf 226}, 289 (1983).
  \bibitem{Wess:1992cp}
  J.~Wess and J.~Bagger,
  Princeton, USA: Univ. Pr. (1992) 259 p.
\bibitem{Kuzenko:2010ef} 
  S.~M.~Kuzenko and S.~J.~Tyler,
  Phys.\ Lett.\ B {\bf 698}, 319 (2011)
  [arXiv:1009.3298 [hep-th]].
\bibitem{Kaku:1978nz} 
  M.~Kaku, P.~K.~Townsend and P.~van Nieuwenhuizen,
  Phys.\ Rev.\ D {\bf 17}, 3179 (1978).
\bibitem{Kaku:1978ea} 
  M.~Kaku and P.~K.~Townsend,
  Phys.\ Lett.\ B {\bf 76}, 54 (1978).
\bibitem{Townsend:1979ki} 
  P.~K.~Townsend and P.~van Nieuwenhuizen,
  Phys.\ Rev.\ D {\bf 19}, 3166 (1979).
\bibitem{Kugo:1982cu} 
  T.~Kugo and S.~Uehara,
  Nucl.\ Phys.\ B {\bf 226}, 49 (1983).
  \bibitem{supergravity}
  D. Z. Freedman and A. Van Proeyen. Supergravity. Cambridge University Press, 2012
\bibitem{Kugo:1982mr} 
  T.~Kugo and S.~Uehara,
  Nucl.\ Phys.\ B {\bf 222}, 125 (1983).
\bibitem{Kallosh:1999jj} 
  R.~Kallosh, L.~Kofman, A.~D.~Linde and A.~Van Proeyen,
  Phys.\ Rev.\ D {\bf 61}, 103503 (2000)
  [hep-th/9907124].
\bibitem{Kallosh:2000ve} 
  R.~Kallosh, L.~Kofman, A.~D.~Linde and A.~Van Proeyen,
  Class.\ Quant.\ Grav.\  {\bf 17}, 4269 (2000)
  [Class.\ Quant.\ Grav.\  {\bf 21}, 5017 (2004)]
  [hep-th/0006179].
\bibitem{Abe:2015nxa} 
  H.~Abe, Y.~Sakamura and Y.~Yamada,
  Phys.\ Rev.\ D {\bf 92}, no. 2, 025017 (2015)
  [arXiv:1504.01221 [hep-th]].
\bibitem{Bergshoeff:2015tra} 
  E.~A.~Bergshoeff, D.~Z.~Freedman, R.~Kallosh and A.~Van Proeyen,
  arXiv:1507.08264 [hep-th].
\end{thebibliography}
\end{document}